\documentclass[aps,prd,twocolumn,showpacs,amsmath,amssymb]{revtex4-1}
\usepackage{amsmath}
\usepackage{graphicx}
\usepackage{subfigure}
\usepackage{epstopdf}
\usepackage{color}
\usepackage{multirow}
\usepackage{setspace}
\usepackage{overpic}
\usepackage{amssymb}
\usepackage[dvipdfmx, bookmarksnumbered, pdfstartview=FitH,colorlinks,urlcolor=blue, citecolor=blue,linkcolor=blue] {hyperref}
\usepackage{lineno}
\usepackage{bm}
\usepackage{rotating}
\usepackage[utf8]{inputenc}
\let\oldequation\equation
\let\oldendequation\endequation

\renewenvironment{equation}
  {\linenomathNonumbers\oldequation}
  {\oldendequation\endlinenomath}

\begin{document}
%\linenumbers

\title{\bf \boldmath
Search for the semileptonic decay $D^{0(+)}\to b_1(1235)^{-(0)} e^+\nu_e$}

\author{
M.~Ablikim$^{1}$, M.~N.~Achasov$^{10,c}$, P.~Adlarson$^{64}$, S. ~Ahmed$^{15}$, M.~Albrecht$^{4}$, R.~Aliberti$^{28}$, A.~Amoroso$^{63A,63C}$, Q.~An$^{60,48}$, ~Anita$^{21}$, X.~H.~Bai$^{54}$
% ...
\\
\vspace{0.2cm}
(BESIII Collaboration)\\
\vspace{0.2cm} {\it
$^{1}$ Institute of High Energy Physics, Beijing 100049, People's Republic of China\\
% ...
$^{4}$ Bochum Ruhr-University, D-44780 Bochum, Germany\\
% ...
$^{10}$ G.I. Budker Institute of Nuclear Physics SB RAS (BINP), Novosibirsk 630090, Russia\\
% ...
$^{15}$ Helmholtz Institute Mainz, Johann-Joachim-Becher-Weg 45, D-55099 Mainz, Germany\\
% ...
$^{21}$ Indian Institute of Technology Madras, Chennai 600036, India\\
% ...
$^{28}$ Johannes Gutenberg University of Mainz, Johann-Joachim-Becher-Weg 45, D-55099 Mainz, Germany\\
% ...
$^{48}$ State Key Laboratory of Particle Detection and Electronics, Beijing 100049, Hefei 230026, People's Republic of China\\
% ...
$^{54}$ University of Jinan, Jinan 250022, People's Republic of China\\
% ...
$^{60}$ University of Science and Technology of China, Hefei 230026, People's Republic of China\\
% ...
$^{63}$ (A)University of Turin, I-10125, Turin, Italy; (B)University of Eastern Piedmont, I-15121, Alessandria, Italy; (C)INFN, I-10125, Turin, Italy\\
$^{64}$ Uppsala University, Box 516, SE-75120 Uppsala, Sweden\\
% ...
$^{c}$ Also at the Novosibirsk State University, Novosibirsk, 630090, Russia\\
% ...
}
}
%%% Local Variables:
%%% mode: latex
%%% TeX-master: "draft_BAM186"
%%% End:

\begin{abstract}
Using $2.93~\mathrm{fb}^{-1}$ of $e^+e^-$ annihilation data collected at a center-of-mass energy $\sqrt{s}=3.773$ GeV with the BESIII detector operating at the BEPCII collider, we search for the semileptonic $D^{0(+)}$ decays into a $b_1(1235)^{-(0)}$ axial-vector meson for the first time.
No significant signal is observed for either charge combination.
The upper limits on the product branching fractions are ${\mathcal B}_{D^0\to b_1(1235)^- e^+\nu_e}\cdot {\mathcal B}_{b_1(1235)^-\to \omega\pi^-}<1.12\times 10^{-4}$ and ${\mathcal B}_{D^+\to b_1(1235)^0 e^+\nu_e}\cdot {\mathcal B}_{b_1(1235)^0\to \omega\pi^0}<1.75\times 10^{-4}$ at the 90\%  confidence level.
\end{abstract}


\maketitle

\oddsidemargin  -0.2cm
\evensidemargin -0.2cm

\section{Introduction}
Semileptonic decays of the $D^{0(+)}$  provide an outstanding platform to explore the dynamics of both weak and strong interactions in the charm sector. The semileptonic $D^{0(+)}$ decays into pseudoscalar and vector mesons have been widely studied in both experiment~\cite{pdg2018} and theory.
Extensive studies of the semileptonic $D^{0(+)}$ decays into axial-vector mesons  $\bar K_1(1270)$ and $b_1(1235)$
play an important role in the understanding of nonperturbative strong-interaction dynamics in weak decays~\cite{isgw,isgw2,khosravi,zuo,cheng,momeni1,momeni2}.
Nevertheless, knowledge of these decays is limited. The observation of the Cabibbo-favored decay  $D^+\to \bar K_1(1270)^0e^+\nu_e$ has been reported by the BESIII experiment ~\cite{bes3-Dp-K1ev}, and evidence for  $D^0\to K_1(1270)^-e^+\nu_e$  has been found at CLEO~\cite{cleo-D0-K1ev}. The measured branching fractions are consistent with theoretical predictions based on the Isgur-Scora-Grinstein-Wise (ISGW) quark model~\cite{isgw} and its upgrade (ISGW2)~\cite{isgw2}, as well as those based on the covariant light-front quark model~\cite{cheng}.
As for the singly Cabibbo-suppressed decays $D^{0(+)}\to b_1(1235)^{-(0)}e^+\nu_e$,
no experimental study has yet been carried out.
Experimental measurements of the semileptonic decays $D^{0(+)}\to b_1(1235)^{-(0)}e^+\nu_e$ are important to test theoretical calculations and to understand nonperturbative effects in heavy meson decays~\cite{isgw,isgw2,cheng}.
Moreover, the observation of the $b_1(1235)^{-(0)}$ meson in semileptonic decays would provide a clean environment to study its nature~\cite{bes3-white-paper}.

In this paper, we report the first search for the semileptonic decays $D^0\to b_1(1235)^- e^+\nu_e$ and $D^+\to b_1(1235)^0 e^+\nu_e$. The data used in this analysis, corresponding to an integrated luminosity of 2.93~fb$^{-1}$ ~\cite{lum_bes3}, was accumulated at a center-of-mass energy of 3.773~GeV with the BESIII detector. Throughout this paper, charge conjugate channels are always implied.

\section{BESIII detector and Monte Carlo simulation}
The BESIII detector is a magnetic
spectrometer~\cite{BESIII} located at the Beijing Electron
Positron Collider (BEPCII)~\cite{Yu:IPAC2016-TUYA01}. The
cylindrical core of the BESIII detector consists of a helium-based
 multilayer drift chamber (MDC), a plastic scintillator time-of-flight
system (TOF), and a CsI(Tl) electromagnetic calorimeter (EMC),
which are all enclosed in a superconducting solenoidal magnet
providing a 1.0~T magnetic field. The solenoid is supported by an
octagonal flux-return yoke with resistive plate counter muon
identifier modules interleaved with steel. The acceptance of
charged particles and photons is 93\% over $4\pi$ solid angle. The
charged-particle momentum resolution at $1~{\rm GeV}/c$ is
$0.5\%$, and the $dE/dx$ resolution is $6\%$ for the electrons
from Bhabha scattering. The EMC measures photon energies with a
resolution of $2.5\%$ ($5\%$) at $1$~GeV in the barrel (end cap)
region. The time resolution of the TOF barrel part is 68~ps, while
that of the end cap part is 110~ps.

Simulated samples produced with the {\sc geant4}-based~\cite{geant4} Monte Carlo (MC) package which
includes the geometric description of the BESIII detector and the
detector response, are used to determine the detection efficiency
and to estimate the backgrounds. The simulation includes the beam
energy spread and initial state radiation (ISR) in the $e^+e^-$
annihilations modeled with the generator {\sc kkmc}~\cite{kkmc}.
The inclusive MC samples consist of the production of $D\bar{D}$
pairs with consideration of quantum coherence for all neutral $D$
modes, the non-$D\bar{D}$ decays of the $\psi(3770)$, the ISR
production of the $J/\psi$ and $\psi(3686)$ states, and the
continuum processes.
The known decay modes are modeled with {\sc
evtgen}~\cite{evtgen} using the branching fractions taken from the
Particle Data Group~\cite{pdg2018}, and the remaining unknown decays
from the charmonium states with {\sc
lundcharm}~\cite{lundcharm}. The final state radiations
from charged final state particles are incorporated with the {\sc
photos} package~\cite{photos}.
The signal process $D^{0(+)}\to b_1(1235)^{-(0)}e^+\nu_e$ is simulated with  $b_1(1235)^{-(0)}$ decaying into $\omega \pi^{-(0)}$, using the ISGW2 model~\cite{isgw2}.
A relativistic Breit-Wigner function is used to parameterize the resonance $b_1(1235)^{-(0)}$, the mass and width of which are fixed to
the world-average values {\color{blue}of} $1229.5\pm 3.2$ MeV{\color{blue}/$c^{2}$} and $142\pm9$ MeV, respectively~\cite{pdg2018}.

\section{Data analysis}

The process $e^+e^-\to \psi(3770)\to D\bar D$ provides an ideal opportunity to study semileptonic $D^{0(+)}$ decays with the double-tag~(DT) method, because there are no additional particles that accompany the $D$ mesons in the final states~\cite{mark3}.
Throughout the paper, $D$ denotes $D^0$ or $D^+$.
At first,
single-tag~(ST) $\bar D^0$ mesons are reconstructed by using the hadronic decay modes of
$\bar D^0\to K^+\pi^-$, $K^+\pi^-\pi^0$, and $K^+\pi^-\pi^-\pi^+$;
while ST $D^-$ mesons are reconstructed via the decays
$D^-\to K^{+}\pi^{-}\pi^{-}$,
$K^0_{S}\pi^{-}$, $K^{+}\pi^{-}\pi^{-}\pi^{0}$, $K^0_{S}\pi^{-}\pi^{0}$, $K^0_{S}\pi^{+}\pi^{-}\pi^{-}$,
and $K^{+}K^{-}\pi^{-}$.
Then the semileptonic $D$
candidates are reconstructed with the remaining tracks and showers.
The candidate event in which $D$ decays into $b_1(1235)e^+\nu_e$ and $\bar D$ decays into a tag mode is
called a DT event.
Since the branching fraction of the subsequent decay $b_1(1235) \to \omega \pi$ is not well measured,
the product of the branching fractions of the decay $D\to b_1(1235)e^+\nu_e$ (${\mathcal B}_{\rm SL}$) and its subsequent decay $b_1(1235)\to\omega\pi$ (${\mathcal B}_{b_1}$) is determined using
\begin{equation}
\label{eq:bf}
{\mathcal B}_{\rm SL}\cdot {\mathcal B}_{b_1} = \frac{N_{\rm DT}}{N^{\rm tot}_{\rm ST}\cdot\bar \varepsilon_{\rm SL}\cdot {\mathcal B}_{\omega} \cdot ({\mathcal B}_{\pi^0})^k},
\end{equation}
where $N_{\rm ST}^{\rm tot}$ and $N_{\rm DT}$ are the yields of the ST $\bar D$ mesons and the DT signal events in data, respectively; ${\mathcal B}_{\omega}$ and ${\mathcal B}_{\pi^0}$ are the branching fractions of  $\omega\to\pi^+\pi^-\pi^0$ and $\pi^0\to \gamma\gamma$, respectively; $k$ is {\color{blue}the component, which corresponds to} the number of $\pi^0$ mesons in the final states and $\bar\varepsilon_{\rm SL}$ is the average efficiency of reconstructing $D\to b_1(1235)e^+\nu_e$.
The average signal efficiency, weighted over the tag modes $i$, is calculated by $\bar\varepsilon_{\rm SL}=\Sigma_i [(\varepsilon^i_{\rm DT}\cdot N^i_{\rm ST})/(\varepsilon^i_{\rm ST}\cdot N^{\rm tot}_{\rm ST})]$, where  $N^i_{\rm ST}$ is the ST yield of $\bar{D}\to i$, $\varepsilon^i_{\rm ST}$ is the detection efficiency of reconstructing $\bar{D}\to i$, and $\varepsilon^i_{\rm DT}$ is the detection efficiency of reconstructing $\bar{D}\to i$ and $D\to b_1(1235)e^+\nu_e$ at the same time.

The ST $\bar D$ candidates are selected with the same criteria employed in our previous works~\cite{epjc76,cpc40,bes3-pimuv,bes3-Dp-K1ev,bes3-etaetapi,bes3-omegamuv,bes3-etamuv,bes3-etaX,bes3-DCS-Dp-K3pi,bes3-D-KKpipi}.
For each charged track (except for those used for reconstructing $K^0_S$ meson decays), the polar angle with respect to the MDC axis ($\theta$) is required to satisfy $|\cos\theta|<0.93$, and the point of closest approach to the interaction point (IP) must be within 1\,cm in the plan perpendicular to the MDC axis and within $\pm$10\,cm along the MDC axis.
Charged tracks are identified by using the $dE/dx$ and TOF information, with which the combined confidence levels under the pion and kaon hypotheses are computed separately. A charged track is assigned as the particle type which has a larger probability.

Candidate $K_S^0$ mesons are formed from pairs of oppositely charged tracks. For these two tracks, the distance of closest approach to the IP is required to be less than 20\,cm along the MDC axis. No requirements on the distance of closest approach in the transverse plane or on particle identification (PID) criteria are applied to these tracks. The two charged tracks are constrained to originate from a common vertex, which is required to be away from the IP by a flight distance of at least twice the vertex resolution. The invariant mass of the $\pi^+\pi^-$ pair is required to be within $(0.486,0.510)$~GeV/$c^2$.

Neutral pion candidates are reconstructed via the $\pi^0\to\gamma\gamma$ decays. Photon candidates are chosen from the EMC showers. The EMC time deviation from the event start time is required to be within [0,\,700]\,ns. The energy deposited in the EMC is required to be greater than 25\,(50)\,MeV if the crystal with the maximum deposited energy in that cluster is in the barrel~(end cap) region~\cite{BESCol}. The opening angle between the photon candidate and the nearest charged track is required to be greater than $10^{\circ}$. For any $\pi^0$ candidate, the invariant mass of the photon pair is required to be within $(0.115,\,0.150)$\,GeV$/c^{2}$. To improve the momentum resolution, a mass-constrained~(1-C) fit to the nominal $\pi^{0}$ mass~\cite{pdg2018} is imposed on the photon pair. The four-momentum of the $\pi^0$ candidate returned by this kinematic fit is used for further analysis.

In the selection of $\bar D^0\to K^+\pi^-$ events, the backgrounds from cosmic rays and Bhabha events are rejected by using the same requirements described in Ref.~\cite{deltakpi}.
To separate the ST $\bar D$ mesons from combinatorial backgrounds, we define the energy difference $\Delta E\equiv E_{\bar D}-E_{\mathrm{beam}}$ and the beam-constrained mass $M_{\rm BC}\equiv\sqrt{E_{\mathrm{beam}}^{2}/c^{4}-|\vec{p}_{\bar D}|^{2}/c^{2}}$, where $E_{\mathrm{beam}}$ is the beam energy, and $E_{\bar D}$ and $\vec{p}_{\bar D}$ are the total energy and momentum of the ST $\bar D$ meson in the $e^+e^-$ center-of-mass frame.
If there is more than one $\bar{D}$ candidate in a specific ST mode, the one with the least $|\Delta E|$ is kept for further analysis.

\begin{figure}[htbp]\centering
\includegraphics[width=1.0\linewidth]{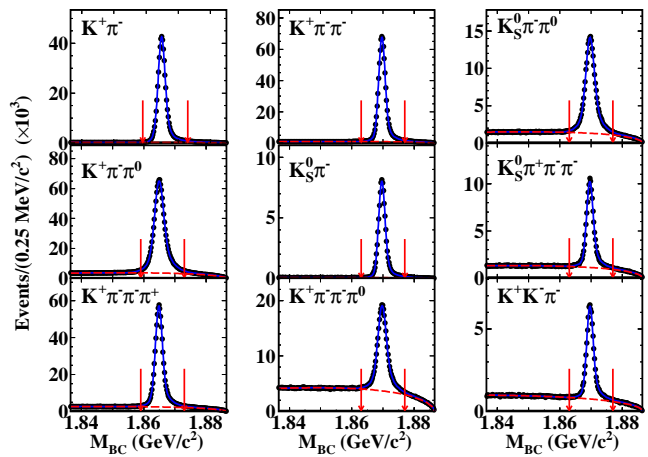}
\caption{
Fits to the $M_{\rm BC}$ distributions of the ST $\bar D$ candidates.
In each plot, the points with error bars are data, the red dashed curve is the background contribution, and the blue solid line shows the total fit.  Pairs of red arrows show the $M_{\rm BC}$ signal windows.}\label{fig:datafit_Massbc}
\end{figure}

To suppress combinatorial backgrounds, the ST $\bar D$ candidates, which are reconstructed by using the modes with and without $\pi^0$ in the final states, are imposed with the requirements of $\Delta E\in (-0.055,0.045)$~GeV and $\Delta E\in (-0.025,0.025)$~GeV, respectively. For each ST mode, the yield of ST $\bar D$ mesons is extracted by fitting the corresponding $M_{\rm BC}$ distribution. The signal is described by an MC-simulated shape convolved with a double-Gaussian function which  compensates the resolution difference between data and MC simulation. The background is parameterized by the ARGUS function~\cite{argus}. All fit parameters are left free in the fits.
Figure~\ref{fig:datafit_Massbc} shows the fits to the $M_{\rm BC}$ distributions for individual ST modes. The candidates with $M_{\rm BC}$ lying in $(1.859,1.873)$ GeV/$c^2$ for $\bar D^0$ tags and $(1.863,1.877)$ GeV/$c^2$ for $D^-$ tags are kept for further analysis. Summing over the tag modes, the total yields of ST $\bar D^0$ and $D^-$ mesons are obtained to be $2321009\pm1875_{\rm stat}$ and $1522474\pm 2215_{\rm stat}$, respectively~\cite{bes3-pimuv}.

We require that there are four and three charged tracks reconstructed in $D^0\to b_1(1235)^- e^+\nu_e$ and $D^+\to b_1(1235)^0 e^+\nu_e$ candidates, respectively. These tracks exclude those used to form the ST $\bar D$ candidates. For each candidate, one charged track is identified as a positron and the others are required to be identified as pions. The selection criteria of charged and neutral pions are the same as those used in selecting the ST $\bar D$ candidates.
To suppress fake $\pi^0$ candidates, the decay angle of $\pi^0$, defined as
\begin{equation}
\cos \theta_{\pi^0}=|E_{\gamma1}-E_{\gamma2}|/|\vec{p}_{\pi^0}\cdot c|, \nonumber
\end{equation}
is required to be less than 0.9. The requirement has been optimized using the inclusive MC sample. $E_{\gamma1}$ and $E_{\gamma2}$ are the energies of the two
daughter photons of the $\pi^0$, and $\vec{p}_{\pi^0}$ is the reconstructed momentum of the $\pi^0$.
For the selected $D^0\to \pi^+\pi^-\pi^-\pi^0e^+\nu_e$ and $D^+\to \pi^+\pi^-\pi^0\pi^0 e^+\nu_e$ candidates,
there are always two possible $\pi^+\pi^-\pi^0$ combinations to form the $\omega$.
The invariant masses of both combinations are required to be greater than 0.6 GeV/$c^2$
to suppress the backgrounds from $D\to a_0(980)e^+\nu_e$.
One candidate is kept for further analysis if either of the combinations has an invariant mass falling in the $\omega$ mass signal region of $(0.757,0.807)$~GeV/$c^2$.
To form a $b_1(1235)$ candidate, the $\omega\pi$ invariant mass is required to be within
$(1.080,1.380)$~GeV/$c^2$.
The background from $D^{0(+)}\to \bar K_1(1270) [K_S^0\pi^{+(0)}\pi^{-(0)}] e^+\nu_e$ is rejected
by requiring the invariant masses of any $\pi^+\pi^-$ ($\pi^0\pi^0$) combinations to be outside
(0.486, 0.510)~GeV/$c^2$~((0.460, 0.510)~GeV/$c^2$).
These requirements correspond to three times the invariant mass resolution about the nominal $K^{0}_{\rm S}$ mass~\cite{pdg2018}.

The $e^+$ candidate is required to have a charge of opposite sign to that of the charm quark in the ST $\bar D$ meson.
The $e^+$ candidate is identified by using the combined $dE/dx$, TOF, and EMC information. The combined confidence levels for the positron, pion, and kaon hypotheses~($CL_e$, $CL_\pi$, and $CL_K$) are computed. The positron candidate is required to satisfy $CL_e>0.001$ and $CL_e/(CL_e+CL_\pi+CL_K)>0.8$. Its deposited energy in the EMC is required to be greater than 0.8 times its momentum reconstructed by the MDC, to further suppress the background from misidentified hadrons and muons.

The peaking backgrounds from hadronic $D$ decays with multiple pions in the final states are rejected by requiring that the invariant mass of  $b_1(1235)e^+$ ~($M_{b_1e^+}$) is less than 1.80~GeV/$c^2$.
To suppress backgrounds with extra photon(s), we require that the energy of any extra photon~($E_{\rm extra}^{\gamma}$) is less than 0.30~GeV and
there is no extra $\pi^0$ ($N_{\rm extra}^{\pi^0}$) in the candidate event.

The neutrino is not detectable in the BESIII detector.  To distinguish semileptonic signal events from backgrounds, we define $U_{\mathrm{miss}}\equiv E_{\mathrm{miss}}-|\vec{p}_{\mathrm{miss}}|\cdot c$, where $E_{\mathrm{miss}}$ and $\vec{p}_{\mathrm{miss}}$
are the missing energy and momentum of the DT event in the $e^+e^-$ center-of-mass frame, respectively.
They are calculated as $E_{\mathrm{miss}}\equiv E_{\mathrm{beam}}-E_{b_1}-E_{e^{+}}$ and $\vec{p}_{\mathrm{miss}}\equiv\vec{p}_{D}-\vec{p}_{b_1}-\vec{p}_{e^{+}}$, where $E_{b_1\,(e^+)}$ and $\vec{p}_{b_1\,(e^+)}$ are the measured energy and momentum of the $b_1(1235)$\,($e^+$) candidates, respectively, and $\vec{p}_{D}\equiv-\hat{p}_{\bar D}\cdot \sqrt{E_{\mathrm{beam}}^{2}/c^{2}-m_{\bar D}^{2}\cdot c^{2} }$, where
$\hat{p}_{\bar D}$ is the unit vector in the momentum direction of the ST $\bar D$ meson and $m_{\bar D}$ is the  nominal $\bar D$ mass~\cite{pdg2018}.
The use of the beam energy and the nominal $D$ mass for the magnitude of the ST $D$ mesons improves the  $U_{\mathrm{miss}}$ resolution. For the correctly reconstructed signal events, $U_{\mathrm{miss}}$  peaks at zero.

\begin{figure*}[htbp]
\includegraphics[width=0.48\linewidth]{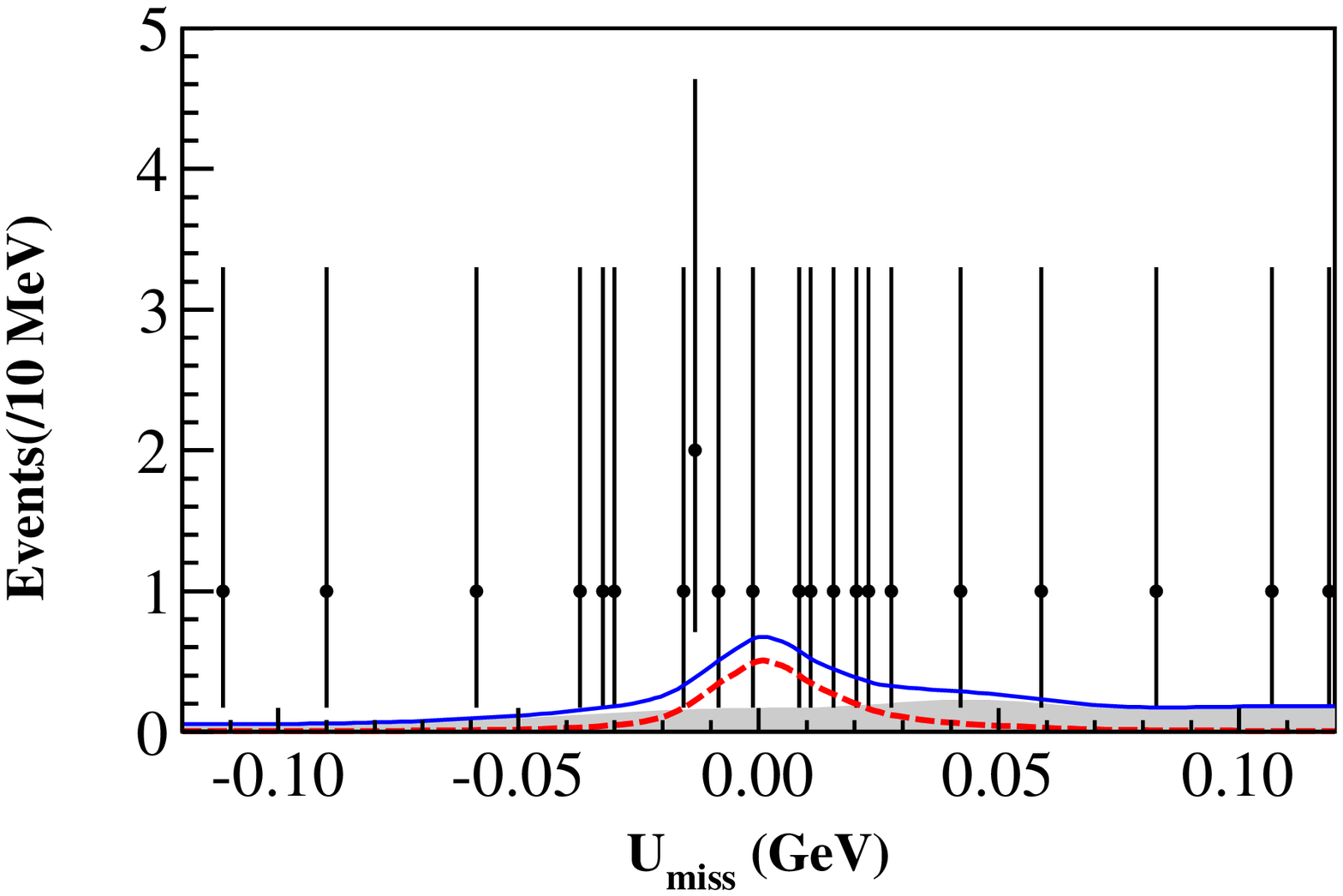}
\includegraphics[width=0.48\linewidth]{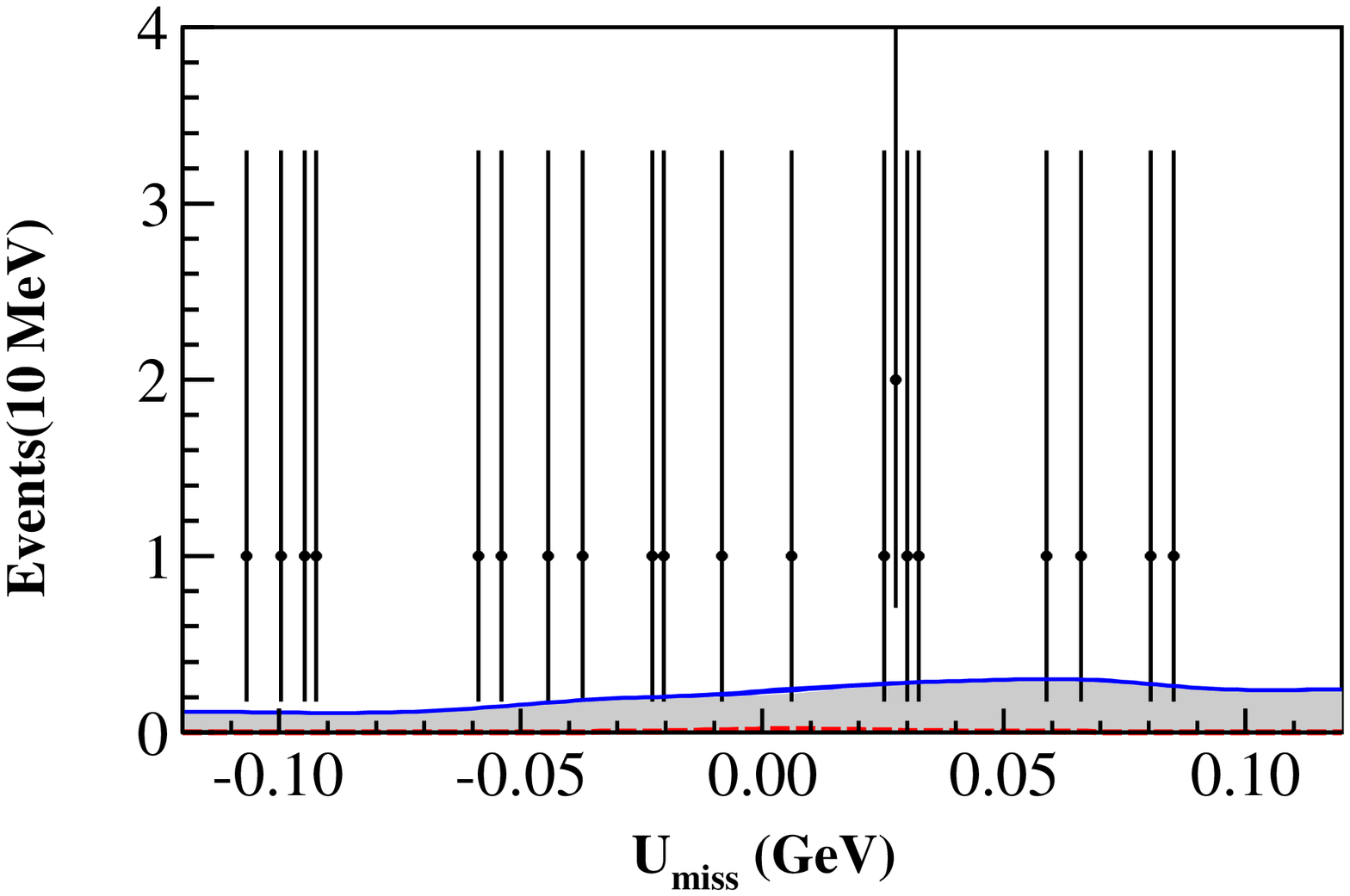}
\caption{Fits to the $U_{\rm miss}$ distributions of the
(left) $D^{0}\to b_1(1235)^{-} e^{+}\nu_e$ and (right) $D^{+}\to b_1(1235)^{0} e^{+}\nu_e$ candidate events.
The points with error bars are data, the red dashed curve is the signal, the gray filled histogram is the background contribution, and the blue solid curve shows the total fit.}
\label{fig:fit_Umistry1}
\end{figure*}

Figure~\ref{fig:fit_Umistry1} shows the $U_{\rm miss}$ distributions of the accepted candidate events.
Unbinned maximum likelihood fits are performed on these distributions. In the fits, the signal and background are modeled by the simulated shapes obtained from the signal MC events and the inclusive MC sample, respectively, and the yields of the signal and background are left free.
Since no significant signal is observed, conservative upper limits will be set by assuming all the fitted signals are from $b_1(1235)$.

The detection efficiencies $\color{blue} \bar \varepsilon_{\rm SL}$ are estimated to be
$0.0704\pm0.0006$ and $0.0412\pm0.0002$ for the $D^0\to b_1(1235)^- e^+\nu_e$ and $D^+\to b_1(1235)^0 e^+\nu_e$ decays, respectively.
The blue dotted curves in Fig.~\ref{fig:prob} show the raw likelihood distributions versus the corresponding product of branching fractions.

\section{Systematic uncertainty}

With the DT method, many systematic uncertainties on the ST side mostly cancel.
The sources of the systematic uncertainties in the measurements of the product of branching fractions are classified into two cases.
The first one is from the uncertainties relying on effective efficiencies and are assigned relative to the measured branching fractions.
The uncertainty associated with the ST yield $N_{\rm ST}^{\rm tot}$ is estimated to be ~0.5\% \cite{epjc76,cpc40,bes3-pimuv}. The uncertainty from the quoted branching fraction of the $\omega\to\pi^+\pi^-\pi^0$ decay is 0.8\%.
The uncertainties from the tracking and PID of $e^\pm$ are studied with a control sample of $e^+e^-\to\gamma e^+e^-$. The uncertainties from the tracking and PID of $\pi^\pm$ and $\pi^0$ reconstruction are obtained by studing a DT control sample $\psi(3770)\to D\bar D$ with hadronic $D$ decays~\cite{epjc76,cpc40}.
The systematic uncertainties from the tracking (PID) efficiencies are assigned as 1.0\% (1.0\%) per $e^\pm$ and 1.0\% (1.0\%) per $\pi^\pm$, respectively.
The $\pi^0$ reconstruction efficiencies include photon finding, the $\pi^0$ mass window, and the 1-C kinematic fit, the systematic uncertainty of which is taken to be 2.0\% per $\pi^0$.
The systematic uncertainty from the $\pi^0$ decay angle requirement is determined to be 2.0\% per $\pi^0$ by studying the DT events of $D^0\to K^-\pi^+\pi^0$ versus $\bar D^0\to K^+\pi^-$ and $K^+\pi^-\pi^-\pi^+$.
The systematic uncertainty associated with the $\omega$ mass window is assigned to be 1.2\% using a control sample of $D^0\to K^0_S\omega$  reconstructed versus the same $\bar D^0$ tags as those used in the nominal analysis.
The systematic uncertainties from the $E_{\rm extra\,\gamma}^{\rm max}$ and $N_{\rm extra,\pi^0}$ requirements are estimated to be 1.4\% and 2.0\% for $D^0\to b_1(1235)^- e^+\nu_e$ and $D^+\to b_1(1235)^0 e^+\nu_e$, respectively, which are estimated using DT samples of $D^0\to K^-e^+\nu_e$ and $D^+\to K^0_Se^+\nu_e$ decays reconstructed versus the same tags as the nominal analysis.
The systematic uncertainty related to the MC generator is estimated using alternative signal MC samples, which are produced by varying the mass and width of the $b_1(1235)$ by $\pm1\sigma$. The maximum changes of the signal efficiencies, 5.1\% and 2.7\%, are assigned as the systematic uncertainties for $D^0\to b_1(1235)^- e^+\nu_e$ and $D^+\to b_1(1235)^0 e^+\nu_e$, respectively. The uncertainties from limited MC statistics, propagated from those of the ST and DT efficiencies, are 0.7\% and 0.9\% for $D^0\to b_1(1235)^- e^+\nu_e$ and $D^+\to b_1(1235)^0 e^+\nu_e$, respectively.
By adding these uncertainties in quadrature, the total systematic errors associated with the signal efficiencies
($\sigma_{\epsilon}$) are obtained to be 8.2\% and 7.3\% for $D^0\to b_1(1235)^- e^+\nu_e$ and $D^+\to b_1(1235)^0 e^+\nu_e$, respectively.

The second kind of systematic uncertainty originates from the fit to the $U_{\rm miss}$ distribution of the semileptonic $D$ decay candidates. It is dominated by the uncertainty from imperfect knowledge of the background shape. The uncertainty associated with the signal shape is negligible. The background shape is obtained from the inclusive MC sample using a kernel estimation method \cite{Cranmer:2000du} implemented in RooFit~\cite{ROOT}. Unlike the other sources of uncertainties, the background shape directly affects the likelihood function. The smoothing parameter of RooKeysPdf is varied within a reasonable range to obtain alternative background shapes.
The absolute change of the signal yield, which gives the largest upper limit on the branching fraction, is taken as the systematic uncertainty ($\sigma_n$).
It is found to be 1.7 for $D^0\to b_1(1235)^- e^+\nu_e$ and 1.1 for $D^+\to b_1(1235)^0 e^+\nu_e$.

\section{Results}

To take into account the first kind of systematic uncertainty in the calculation of the upper limits, the raw likelihood distribution versus the product of branching fractions is smeared by a Gaussian function with a mean of 0 and a width equal to $\sigma_{\epsilon}$ according to Refs.~\cite{K.Stenson:2006,cpc:up}.

 To incorporate the second kind of systematic uncertainty, the updated likelihood is then convolved with another Gaussian function with mean of 0 and a width equal to $\sigma_{\mathcal B}$ similarly. Here $\sigma_{\mathcal B}$ is an uncertainty of the product of the branching fractions calculated with Eq.~(\ref{eq:bf}) by replacing $N_{\rm DT}$ with $\sigma_n$.

\begin{figure*}[htbp]
\includegraphics[width=0.48\linewidth]{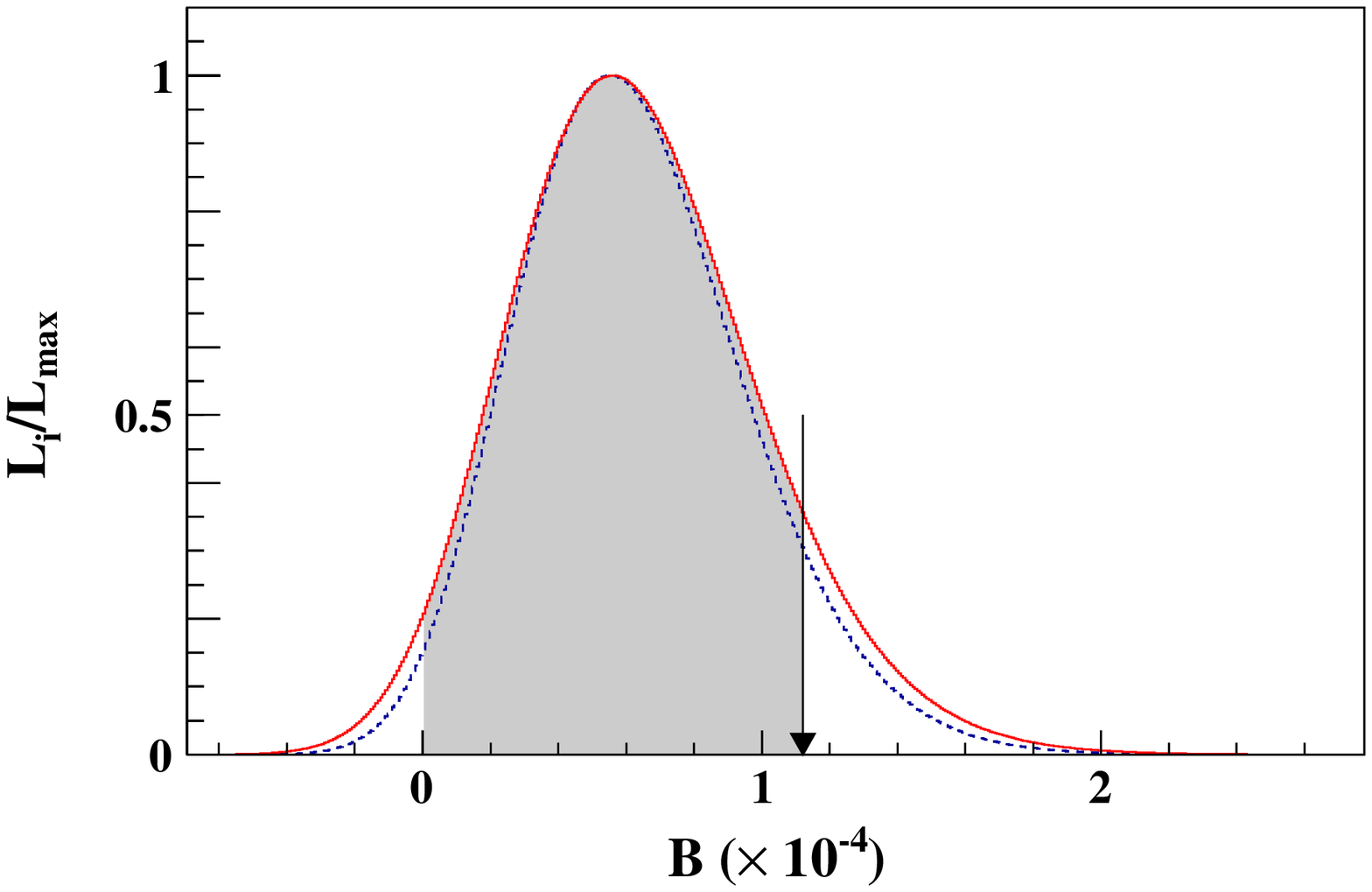}
\includegraphics[width=0.48\linewidth]{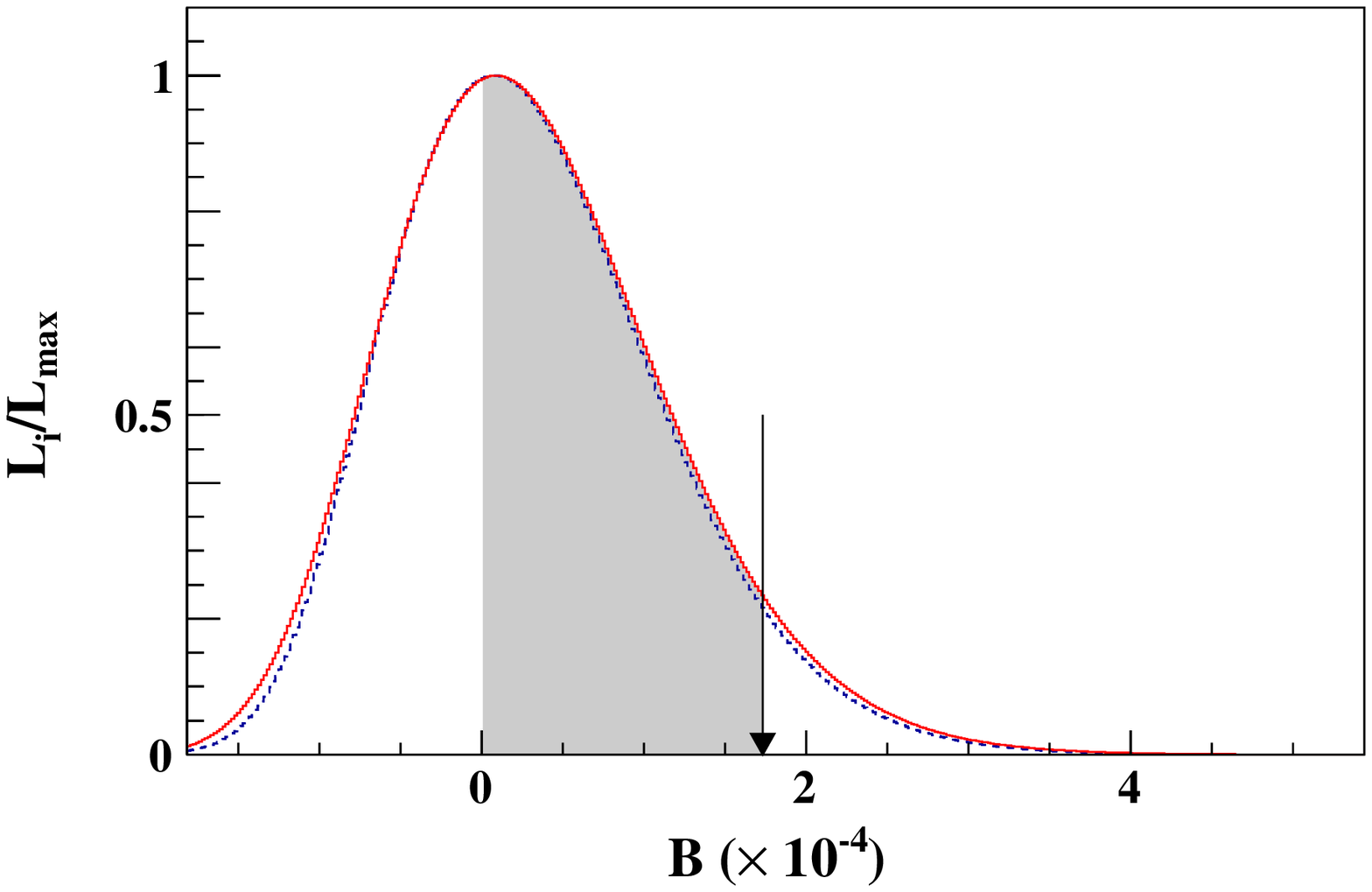}
\caption{Likelihood distributions versus the corresponding product of branching fractions for
(left) $D^{0}\to b_1(1235)^{-} e^{+}\nu_e$ and (right) $D^{+}\to b_1(1235)^{0} e^{+}\nu_e$, with (red solid curves) and without (blue dotted curves) smearing the systematic uncertainties.
The black arrows correspond to the upper limits at the 90\% confidence level.}
\label{fig:prob}
\end{figure*}

The red solid curves in Fig.~\ref{fig:prob} show the resulting likelihood distributions for the two decays.
The upper limits on the product of branching fractions at the 90\% confidence level (C.L.), obtained
by integrating $L(\mathcal B)$ from zero to 90\% of the total curve, are
\begin{eqnarray}
{\mathcal B}_{D^0\to b_1(1235)^- e^+\nu_e}\cdot {\mathcal B}_{b_1(1235)^-\to \omega\pi^-}< 1.12\times 10^{-4} \nonumber
\end{eqnarray}
and
\begin{eqnarray}
{\mathcal B}_{D^+\to b_1(1235)^0 e^+\nu_e}\cdot {\mathcal B}_{b_1(1235)^0\to \omega\pi^0}<1.75\times 10^{-4}. \nonumber
\end{eqnarray}

\section{Summary}

In summary, by analyzing $2.93~\mathrm{fb}^{-1}$  of $e^+e^-$ collision data taken at $\sqrt{s}=3.773$~GeV with the BESIII detector, the semileptonic $D^{0(+)}$ decays into axial-vector mesons $b_1(1235)^{-(0)}$ have been searched for for the first time. Since no significant signal is observed, the upper limits on the product of branching fractions for $D^0\to b_1(1235)^- e^+\nu_e$ and $D^+\to b_1(1235)^0 e^+\nu_e$ at the 90\% C.L. are estimated to be ${\mathcal B}_{D^0\to b_1(1235)^- e^+\nu_e}\cdot {\mathcal B}_{b_1(1235)^-\to \omega\pi^-}<1.12\times 10^{-4}$ and ${\mathcal B}_{D^+\to b_1(1235)^0 e^+\nu_e}\cdot {\mathcal B}_{b_1(1235)^0\to \omega\pi^0}<1.75\times 10^{-4}$, respectively. When assuming ${\mathcal B}_{b_1(1235)\to \omega\pi}=1$, these results are comparable with the theoretical prediction in Ref.~\cite{cheng}. It is anticipated that these decays could be observed with larger data samples at BESIII~\cite{bes3-white-paper} and Belle~II~\cite{belle2-white-paper}.

\section{Acknowledgement}

The BESIII collaboration thanks the staff of BEPCII and the IHEP computing center for their strong support. This work is supported in part by National Key Basic Research Program of China under Contract No. 2015CB856700; National Natural Science Foundation of China (NSFC) under Contracts Nos. 11805037, 11625523, 11635010, 11735014, 11822506, 11835012, 11935015, 11935016, 11935018, 11961141012; the Chinese Academy of Sciences (CAS) Large-Scale Scientific Facility Program; Joint Large-Scale Scientific Facility Funds of the NSFC and CAS under Contracts Nos. U1832121, U1732263, U1832207; CAS Key Research Program of Frontier Sciences under Contracts Nos. QYZDJ-SSW-SLH003, QYZDJ-SSW-SLH040; 100 Talents Program of CAS; INPAC and Shanghai Key Laboratory for Particle Physics and Cosmology; ERC under Contract No. 758462; German Research Foundation DFG under Contracts Nos. 443159800, Collaborative Research Center CRC 1044, FOR 2359, FOR 2359, GRK 214; Istituto Nazionale di Fisica Nucleare, Italy; Ministry of Development of Turkey under Contract No. DPT2006K-120470; National Science and Technology fund; Olle Engkvist Foundation under Contract No. 200-0605; STFC (United Kingdom); The Knut and Alice Wallenberg Foundation (Sweden) under Contract No. 2016.0157; The Royal Society, UK under Contracts Nos. DH140054, DH160214; The Swedish Research Council; U. S. Department of Energy under Contracts Nos. DE-FG02-05ER41374, DE-SC-0012069.

\end{document}